\documentstyle[aps,psfig]{revtex}

\newcommand{\eqn}[1]{eqn.~(\ref{eq:#1})}
\newcommand{\fig}[1]{fig.~\ref{fig:#1}}
\newcommand{\lbl}[1]{\label{eq:#1}}
\def\br{{\bf r}}
\def\<{\langle}
\def\>{\rangle}
\title{Transforming Signs to Phase Distributions in Quantum
Simulations.}
\author{J.M. Deutsch\\
	University of California, Santa Cruz \\
	Santa Cruz, CA 95064}
\begin{document}
\maketitle
\begin{abstract}
A method is developed which speeds up averaging in quantum
simulations where minus signs cause difficulties.
A Langevin equation
method in conjunction with a replication algorithm is used
enabling one to average over a continuously varying complex
number. Instead of ensemble averaging this number directly,
the phase of the complex number is followed over time.
The method is illustrated in some simple cases where the answers
obtained can be compared to exact results, and also compared to
conventional averaging procedures which converge orders of magnitude
slower than this method. Limitations of this method are also
described.
\end{abstract}

\section{Introduction}
\label{introduction}
Simulation of quantum many body systems have been greatly hindered
by the ``sign problem". Quantum systems with many degrees of
freedom are often simulated using stochastic methods. For a variety
of problems, usually involving fermions, an average over many negative
and positive numbers must be taken. Sometimes the sign does
not effect results~\cite{sorella2} but in many interesting cases
the resulting average is
very small because of cancellations, so that in order
to obtain good statistics, the number
of realizations that must be taken is exponential in the inverse
temperature,
making it infeasible to get information about ground state properties
of a system.

The main idea of this work can be understood by considering the
following example. It will be seen below that it describes the
quantum mechanics of $n$ bosons restricted to one site. Consider
the stochastic equation for $z(t)$ which is a complex function of
time:
\begin{equation}
\dot z = i~f(t)z
\lbl{example}
\end{equation}
Here $f(t)$ is real gaussian noise with correlation function
$\langle f(t)f(t')\rangle ~=~ v\delta (t-t')$, where the
angled brackets denote an average over the noise $f(t)$. Choosing
$z(0) = 1$ the above equation is easily solved giving
\begin{equation}
z = e^{i\theta t}
\end{equation}
where
\begin{equation}
\theta(t) ~=~ \int_0^t f(\tau ) d\tau
\lbl{randomwalk}
\end{equation}
Here $\theta (t)$ describes a random walk in time, and therefore
the the probability distribution of $\theta$ at time $t$, $P(\theta ,
t)$,
is a gaussian distribution
with zero mean and variance $\sigma ^2 ~=~ vt$.
Below we will see that the ground state
energy of $n$ bosons is simply obtained from the asymptotic
behavior of $\langle z(t)^n\rangle$, which is solved for as follows:
\begin{equation}
\langle z(t)^n\rangle ~=~ \langle e^{i n \theta}\rangle
~=~\int P(\theta,t) e^{in\theta}d\theta ~=~ e^{-n^2 v t/2}.
\end{equation}

Now let us turn to the problem of how, without the aid
of this analytical solution,
one could obtain $\langle z(t)^n\rangle$ numerically. Because
$|z(t)| ~=~ 1$, to obtain the correct average would
require over $e^{n^2 v t}$ independent runs of \eqn{example}
to obtain adequate statistics. This can be very large.

The alternative to this direct averaging procedure is to obtain
the probability distribution $P(\theta , t)$. In this case it is
much better to keep track of the total angle
$\theta$, not $\theta ~{\rm mod}~ (2\pi )$. Keeping track of
the total angle is done as \eqn{example} is evolved. The increment in
$\theta$ found in a single time step is added to the previous value
of $\theta$ and
a histogram of $\theta$ at a given time can be obtained by running
\eqn{example} over many realizations of $f$.

Does analytically continuing $\theta$ and then
finding $P(\theta, t)$ help in obtaining $\langle z(t)^n\rangle$?
In this case and in the less trivial cases considered below,
it appears to reduce the computation time from something exponential
to algebraic. Of course assumptions must be made in order to obtain
such a
reduction, the most important one being that $P(\theta, t)$
is a smooth function of $\theta$. There are also problems encountered
in
deciding what function to use in fitting $P$, as will be discussed
later.
In the case of interest here, fitting $\ln P(\theta, t)$ to
a quadratic function of $\theta$ gives accurate results.
In fact the ground state energies should be obtained to within
an error $N^{-1/2}$ where $N$ is the number of independent runs.

Having motivated the method that will be considered, the Langevin
equation that simulates quantum many body systems will be described
and numerical results will be given.

\section{Simulation Method}

A system of fermions or bosons can be simulated by considering
a stochastic equation which is a generalization of \eqn{example}.
A field $\phi (\br ,t)$ is governed by the equation
\begin{equation}
\dot\phi (\br ,t) = f(\br ,t) \phi + {1\over 2m} \nabla^2 \phi
\lbl{langevin}
\end{equation}
where $\langle f(\br ,t) f(\br ',t)\rangle = \delta (t-t') v(\br -\br
')$.
This describes a system of bosons of mass $m$
interacting with potential
\begin{equation}
U ~=~ -\sum_{i<j}^n v(\br_i-\br_j) - {n\over 2} v(0)
\lbl{potential}
\end{equation}
evolving in imaginary time.
The wave function at time $t$, is
\begin{equation}
\Psi (\br_1,\dots,\br_n) ~=~
\langle \phi(\br_1)\dots\phi(\br_n)\rangle.
\lbl{wavefunction}
\end{equation}

To simulate $n$ fermions, $n$ fields $\phi_1,\phi_2,\dots ,\phi_n$ are
evolved starting from different initial conditions but with the
same $f(\br ,t)$. The wave function at time $t$
in this case can be expressed
in terms of the average of a Slater determinant:
\begin{eqnarray}
\Psi(\br_1,\dots,\br_n) \:  = \langle
\: \left|
\begin{array}{ccc}
\phi_1(\br_1) & \dots & \phi_1(\br_n) \\
 . &   & . \\
 . &   & . \\
 . &   & . \\
\phi_n(\br_1) & \dots & \phi_n(\br_n)
\end{array}
\; \; \right| \rangle
\lbl{determinant}
\end{eqnarray}

Note that the potential in \eqn{potential} is the {\em negative}
of the correlation function. Therefore to simulate repulsive potentials
requires an $f(\br ,t)$ that is complex~\cite{white}.
This method is closely related to
a simulation method for the density matrix~\cite{koonin,sorella1}
and also to projector Monte-Carlo~\cite{blankenbecler}.

In order to obtain the ground state energy for quantum systems such
as the ones described above requires determining the asymptotic
exponential decay of the wave function. In principal this can be
done by averaging over many realizations of $f$, however
there are two reasons why this is impractical.

The first is a problem of importance sampling. As an example,
consider determining the ground state energy $E(n)$ of n bosons. This
is obtained by calculating $<\phi^n(\br,t)>$ and fitting this
to an exponential $\exp(E(n)t)$ for long times. The problem is that
$E(1) \neq E(n)/n$. As a result the typical behavior of $\phi^n(\br,t)$
and the average behavior are very different. What dominates the average
are some very rare realizations of $f$. Another way of describing
such behavior is in terms of the notion of
intermittency~\cite{paladin}.
In order to get the correct average, one must use importance sampling.
There are two ways to do this. One is to use Monte-Carlo keeping
track of all time paths~\cite{koonin}. The second is to use a
replication
algorithm~\cite{ceperley,kung}. In order to keep track of
angles, it is much more convenient to use the second method.

A large number of copies of systems described by \eqn{langevin}
are run in parallel. The number of configurations that are replicated
is proportional to their weight. For example, if we wish to measure
$\<|\phi|^{n+1}\>/\<|\phi|^n\>$ at some
site $\br$, we should choose a weight proportional to $|\phi|^n$ .
The proportionality
factor is chosen so that the number of copies of the system
stays almost constant. This method is similar to others
described in detail in refs. ~\cite{ceperley} and ~\cite{kung}.

The second problem is one encountered when simulating repulsive bosons
and fermions. Here non-positivity leads to the problem
discussed in the introduction, that
an exponentially large number of runs
must be taken to obtain adequate statistics. The
phase method proposed here is designed to reduce the number
of runs. This
will be illustrated numerically for a system of repulsive bosons. One
should note that although it is possible to choose a method
for this problem
where there are no minus signs~\cite{hirsch,suzuki}, this is an
instructive
example. It will be seen that the method used generalizes
to fermions.

\section{Obtaining the phase}
Consider $n$ bosons on a  one dimensional lattice with $L$
sites interacting via an on-site repulsive interaction $U$.
This has been the focus of recent investigation~\cite{batrouni} using
``world line" Monte Carlo~\cite{hirsch,suzuki}, which is clearly better
suited
to this problem than the method here. We use the method
above only to illustrate techniques to cope with negative
signs.
In obtaining ground state energies, any component of $|\Psi>$ can
be measured.
As will be seen below, in order to make the phase method work
it is important to make a judicious choice for this.
We start by choosing to measure $\Psi (x_1,x_2,\dots,x_n)$ where
the $x_i$'s are all different sites. Later we will see why
using the same sites does not work as well. The prescription
for computing $\Psi$ is given by \eqn{wavefunction}.

The replication method outlined above was the case where
all the $x_i$'s are the same. To generalize the discussion
above, one just
chooses a weight $w \equiv |\phi(x_1)\phi(x_2)\dots\phi(x_n)|$.
The ground state energy of $n+1$ particles can be easily computed
by knowing
\begin{equation}
{<\phi(x_1)\dots\phi(x_{n+1})>\over <|\phi(x_1)\dots\phi(x_n)|>}
~=~
{<e^{i\theta}|\phi(x_{n+1})|>_w}
~=~ \int P(\Theta ,t) e^{i\Theta} d\Theta
\lbl{transform}
\end{equation}
where the last average is done with respect to the weight $w$,
and $\theta$ is the total phase angle of
$\phi(x_1)\dots\phi(x_{n+1})$.
$P(\Theta ,t) \equiv \<|\phi_{n+1}|\delta(\Theta - \theta(t))\>_w$
is closely related to $P(\theta , t)$ described below
\eqn{randomwalk}.

\begin{figure}[tbh]
   \begin{center}
   \
   \psfig{file=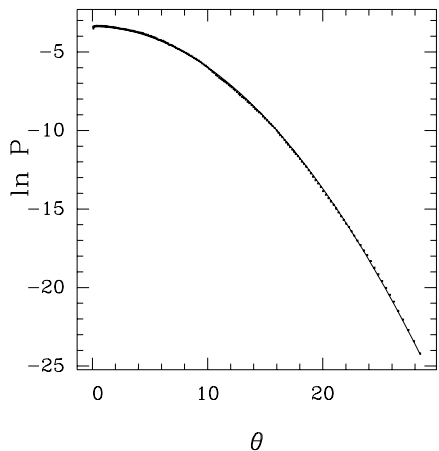,height=2.5in}
   \end{center}
\caption[Histogram of $\Theta$ ] {The histogram
$P(\theta , t)$ at $t=5$. The histogram was obtained
by using the replication algorithm in conjunction with \eqn{langevin}
.
The algorithm was run by giving favorable weights to large angles
enabling good statistics for large values of the angle to be obtained.
The
solid line is a quadratic fit to the data.}
\label{fig:P.05}
\end{figure}

\begin{figure}[tbh]
   \begin{center}
   \
   \psfig{file=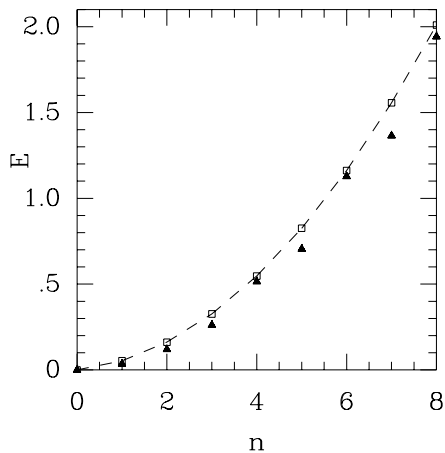,height=2.5in}
   \end{center}
\caption[Energies for U=0.5 ] {The ground state energy $E$,
as a function of number of particles n,
for repulsive bosons with interaction energy of 0.5.
The solid triangles are the results obtained using $P(\Theta,t)$.
The open squares with the dashed line going through them are the
exact results.}
\label{fig:diffsites}
\end{figure}

To test this out numerically, the number of lattice sites
was chosen to be 8, and we started out by using a small
value for  $U ~=~ 0.5$. An average of 4497 copies were replicated.
From this data $P(\Theta ,t)$ was obtained with good statistics.
In order to obtain an improved estimate of the tails of this
distribution, the replication algorithm  was altered to weight
large $\Theta$, that is by letting $w \rightarrow
w\exp(const.\Theta)$.
By changing the constant in the exponent, $P(\Theta ,t)$
was probed for different
regions of $\Theta$. By combining these, one obtains a good estimate
for $P(\Theta ,t)$ over nine orders of magnitude. An example of
$\ln P(\Theta ,t)$ is shown in \fig{P.05} for $n=8$. The solid line
is a quadratic fit to the data. By taking the fourier transform
of this for different times, as prescribed by \eqn{transform}, the
ground state
energy can be obtained. This is shown in \fig{diffsites} by the
solid triangles. The open squares show the exact results for
comparison.

\begin{figure}[tbh]
   \begin{center}
   \
   \psfig{file=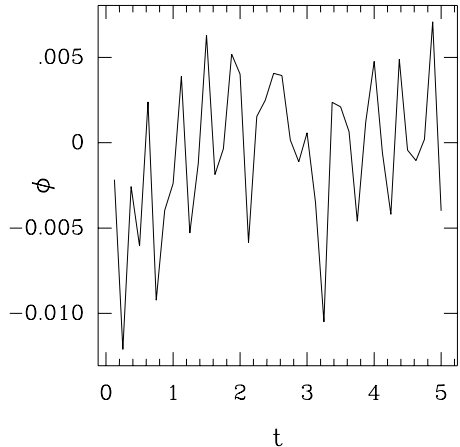,height=2.5in}
   \end{center}
\caption[Conventional averaging]{The result of averaging the
field over the same runs as was used to obtain \fig{P.05}.
An exponential decay is not discernable and many more
averages would be needed to see it. The real part is plotted here.}
\label{fig:phi}
\end{figure}

Without using $P(\Theta ,t)$, one can compute  the weighted average
of $|\phi(x_{n+1})|\exp(i\Theta)$ directly. For comparison
the average of this, over the same copies, is shown \fig{phi}.
It is clear that results obtained by direct averaging are far
inferior to those obtained using $P(\Theta ,t)$.

\section{Improvements to the method}

The method as it stands has difficulties for large $U$. The
histogram of $\Theta$ has too large a variance. This can be
understood by considering a variant of \eqn{example} in
which there are
two independent random processes with different
$f$'s,
$\<f_1(t)f_1(t')\> ~=~ v_1\delta(t-t')$, and
$\<f_2(t)f_2(t')\> ~=~ v_2\delta(t-t')$. The equations
corresponding to each are $\dot z_i ~=~ if_i(t) z_i$.
and therefore
$z_i ~=~ e^{i\theta_i(t)}$, $i=1,2$ .
Suppose we are interested in numerically computing the sum
\begin{equation}
\<A_1 z_1^n + A_2 z_2^n\> = A_1 e^{-n^2v_1t/2}+A_2 e^{-n^2v_2t/2}
\lbl{average}
\end{equation}
This average will be dominated
for long times by the $z_i$ with the smaller of the $v_i$'s, say
$v_1$.
Under what
circumstances will the numerical method of finding the histogram
for the phase angle of the above sum give a sensible answer?
Eqn. (\ref{eq:average}) can be thought of
as the sum of two randomly rotating vectors with magnitudes $|A_i|$.
From this observation, it is easy to see that for long times,
the variance of the phase
angle will be the same as that for $z_1^n$ if $|A_1| > |A_2|$.
For $|A_1| < |A_2|$ a broader
distribution will be obtained. Unless very careful fitting
to a non-quadratic expression is done, an erroneously high
energy will be obtained.

\begin{figure}[tbh]
   \begin{center}
   \
   \psfig{file=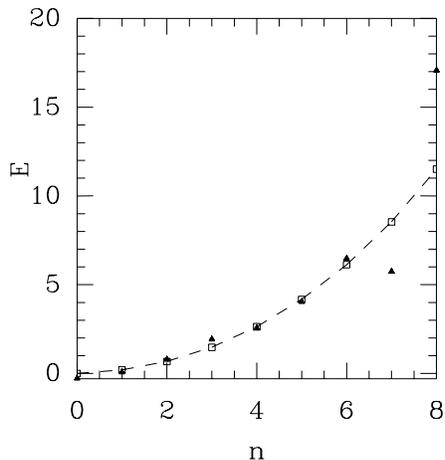,height=2.5in}
   \end{center}
\caption[Filtered data]{The ground state energy $E$,
as a function of number of particles n,
for repulsive bosons with interaction energy of 4.0,
obtained using linear filtering of the field (solid triangles),
and the exact energy (open squares with the dashed line going through
them).
}
\label{fig:filter}
\end{figure}

This example shows that very careful fitting is necessary
if the wave function has too high an amplitude of excited
states. One way to reduce this effect is to choose to measure
$\Psi$ on different sites as was done in the numerical
example above. Another way to improve this method is to
{\em filter} $\phi$ in time. It is easily shown that applying
a linear filter to $\phi (t)$, $\gamma(t) \equiv \phi(t) * g(t)$
does not change the ground state
energy if the filtering function $g(t)$ decays sufficiently rapidly.
For example one can evolve the equation
\begin{equation}
\dot \gamma(\br,t) ~=~ -c\gamma(\br,t) + \phi(\br,t)
\end{equation}
simultaneously with evolving \eqn{langevin} and measure $\gamma^n$
instead
of $\phi^n$. This method
works quite well. The energies for an eight site lattice
for the case $U~=~ 4$ are shown in \fig{filter}. The energies
are in good agreement with the exact results up to $n ~=~ 6$
after which it becomes apparent that the higher energy states
start to dominate the answer. The method
will presumably improve if in addition to filtering, different
sites are used as in the example above.
   \begin{figure}[t*]
   \begin{center}
   \
   \psfig{file=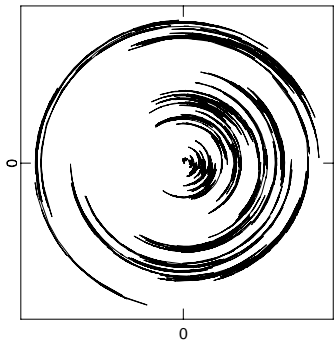,height=2.5in}
   \end{center}
\caption[Log polar plot]{A log polar plot of the Slater
determinant defined in \eqn{determinant} as a function
of time, for three particles on a lattice of 4 sites
with a repulsive potential of 2.5.}
\label{fig:theta}
\end{figure}

Lastly, the case of fermions has been considered. The determinant
formulation of this problem can be used to obtain a continuous
phase as a function of time. Fig. (\ref{fig:theta}) shows
the complex determinant $d$ on a log polar plot, that is
with radius $-\ln |d(t)|$ and angle $arg(d(t))$, for 3 particles
on a 4 site chain, with $U = 2.5$. The phase randomly rotates,
similar to the case of bosons.

In conclusion, a new method for dealing with cancellations
in quantum systems has been developed and tried out in
some simple cases. Further theoretical development of
the determination of $P(\Theta ,t)$ could greatly improve
fitting. Recent theoretical work~\cite{samson} relating the phase used
here to the Berry phase for smooth paths should be further
explored in the context of this simulation method.
It would also be interesting to consider
more difficult problems such as the two dimensional Hubbard
model with repulsive interactions using this method.

\section*{Acknowledgment}

The author thanks Eliot Dresselhaus, Richard Scalettar, and Peter Young
for
useful discussions. This research
was supported by the NSF under grant DMR-9112767.


\end{document}